\documentclass[%
 reprint,twocolumn,
 amsmath,amssymb,
 aps,
]{revtex4-2}

\usepackage{graphicx}
\usepackage{dcolumn}
\usepackage{bm}
\usepackage{epstopdf}
\usepackage{multirow}
\usepackage{braket}
\usepackage{tcolorbox}
\usepackage[breaklinks=true,colorlinks]{hyperref}
\usepackage{multirow}
\usepackage{array}
\usepackage{hyperref}
\usepackage{makecell}

\begin{document}

\preprint{APS/123-QED}

\title{
Generation of large Fock  states from coherent states using Kerr interaction and displacement 
}
\author{Nilakantha Meher}
\email{nilakantha.meher6@gmail.com}
\affiliation{Department of Physics, SRM University-AP, Amaravati, Andhra Pradesh 522240, India}

\author{Anirban Pathak}
\email{anirban.pathak@jiit.ac.in}
\affiliation{Department of Physics and Materials Science and Engineering, Jaypee Institute of Information Technology, A 10, Sector 62, Noida, UP, 201309, India}

\author{S. Sivakumar}
\email{sivakumar.srinivasan@krea.edu.in}

\affiliation{ Krea University, Andhra Pradesh 517646, India}

\begin{abstract}

    We discuss a scheme to generate large Fock states.  The scheme involves repeatedly applying an experimentally feasible unitary transformation to convert a semiclassical state into a Fock state.   The transformation combines  Kerr interaction,  which is a non-Gaussian operation, and pulsed coherent drives.   We identify suitable parameter values (Kerr strength, pulse timings, displacement amplitude) for the physical processes to implement the transformation and generate large Fock states with near-unity fidelity.   The feasibility of implementing the scheme in circuit QED architectures is discussed. The method is also suitable for generating Fock states of cavity fields.
\end{abstract}

\maketitle

\section{Introduction}

The electromagnetic field is the ubiquitous information carrier in all conventional communications.   The quantum properties of the field are rarely  consequential in that context.  However, 
the quantum states of the electromagnetic field possess features that make them suitable for implementing quantum information processing protocols \cite{Braunstein2005RevModPhys,pathak2013elements}. 
To that end,  there is considerable interest in the generation \cite{Stoler1986PRA, Yamamoto1986RevModPhys, Ourjoumtsev2011Nature, rosiek2024quadrature} of various nonclassical states that have a range of applications in quantum information processing \cite{Meher2022EPJP, Northup2014NatPhy}, communication \cite{Gisin2007NatPhot}, quantum computation \cite{Yoran2003PRL}, quantum simulation \cite{sturges2021quantum}, quantum metrology \cite{CavesPRD1981}, etc.   Among the nonclassical states, Fock states, which are also the most nonclassical ones \cite{mandel1986non},  are very relevant to many of the aforementioned applications \cite{Yamamoto1986RevModPhys,Monroe2002Nature,Yoran2003PRL,Santos2005PRL,sturges2021quantum}. They provide a fundamental building block for quantum information processing tasks \cite{Gisin2007NatPhot,Knill2001Nature}, serving as qubits with a definite photon number. Their precise photon number enables robust manipulation and control, facilitating entanglement generation \cite{Malik2016NatPhot,Cable2007PRL} and quantum gate implementation \cite{Santos2005PRL,yanagimoto2020engineering}. Thus, generation of Fock states has been a subject of active research and there have been many experimental techniques \cite{McKeever2004Science, Varcoe2000Nature, Bertet2002PRL, Hofheinz2008Nat, Premaratne2017NatComm} and theoretical proposals \cite{PhysRevLett.115.163603,Uria2020PRL,Domokos1994PRA, Krause1987PRA} to realize these states in various physical systems. 
      
The deterministic generation of Fock states in quantum optical systems can be achieved by evolving the vacuum state or a known initial state in a controlled way through nonlinear interactions \cite{kilin1995fock,Lingenfelter2021ScAdv,Leonski1994PRA,gevorgyan2012generation,Uria2020PRL,Premaratne2017NatComm, Miranowicz2013PRA,Domokos1994PRA, rivera2023creating}.  In cavity QED systems, Fock states are generated by evolving the cavity field state into a Fock state through controlled atom-cavity interactions \cite{Hofheinz2008Nat,Premaratne2017NatComm,PhysRevLett.115.163603,Uria2020PRL,Domokos1994PRA} or by sequentially sending excited atoms through the cavity, where each atom, interacting with the cavity field, emits a photon, thus building up a Fock state \cite{Varcoe2000Nature, Bertet2002PRL, Sayrin2011Nature}. These techniques have enabled the high-fidelity generation of low-photon-number Fock states up to $n\leq 3$, while the high fidelity generation of higher-order Fock states is experimentally challenging \cite{Wang2008prl, Premaratne2017NatComm,Sayrin2011Nature}. Recently, a theoretical scheme has been proposed to generate Fock states containing up to 100 photons in an atom-cavity system \cite{Uria2020PRL}. However, the generated states are not exactly Fock states and are mixed even without dissipation.


Another mechanism for generating Fock states is a driven cavity with a Kerr nonlinear medium, where Fock states are realized through photon blockade \cite{Imamoglu1997PRL, Fushman2008Science, Birnbaum2005Nature,Lingenfelter2021ScAdv,Miranowicz2013PRA}. The anharmonicity in the energy levels of the cavity field, arising due to Kerr nonlinearity,  enables a specific number of photons to be generated in the cavity if the driving frequency is resonant with one of the several transitions \cite{Birnbaum2005Nature, Andrew2021ScAdv, Miranowicz2013PRA,Leonski1994PRA}.  However, generating large Fock states requires a giant Kerr nonlinearity to achieve sufficient anharmonicity in energy levels. With the current experimentally achievable nonlinear strengths, the generation of lower-order Fock states has been demonstrated for photon numbers up to $n\leq 3$ \cite{chakram2022multimode}. Here, we discuss Fock state generation from a coherent state in a nonlinear cavity, i.e., a cavity filled with a nonlinear medium.  The transformation of the initial coherent state to a Fock state is realized by repeated external driving that is linearly coupled to the cavity. We consider Kerr nonlinearity, as it is experimentally realizable. Importantly, this method does not require a large  Kerr nonlinearity.    Mathematically, the process is represented as the repeated application of  successive  unitary  evolutions under $\hat U_{K}(\chi)=e^{-i\chi \hat a^{\dagger 2}\hat a^2}$ corresponding to the Kerr interaction of strength $\chi$ and  the displacement operator $\hat D(\beta)=e^{\beta \hat a^\dagger-\beta^* \hat a}$ on an initial coherent-state $\ket{\alpha}$, where $\beta$ is the displacement amplitude and $\alpha$ is the amplitude of the coherent state.  Such repeated operations of combined $\hat U_{K}(\chi)$ and $\hat D(\beta)$  have been considered earlier, though only for generating a single-photon Fock state starting from the vacuum state \cite{Leonski1994PRA,gevorgyan2012generation}. The nonlinear Kerr unitary operation $\hat U_{K}(\chi)$ is a non-Gaussian operation \cite{Lloyd1999PRL} and has been of interest in various applications \cite{Combes2018PRA, Kitagawa1986PRA, Sundar1996PRA}.   The action of  $\hat U_K(\chi)$ or $\hat D(\beta)$ on a coherent state $\ket\alpha$,  resulting in  $\hat U_{K}(\chi)\ket{\alpha}$ and $\hat D(\beta)\ket{\alpha}$ respectively,    does not alter the photon statistics (photon number fluctuation). However, the combined action represented by $\hat{U}(\beta,\chi)=\hat D(\beta)\hat U_{K}(\chi)\ket{\alpha}$ does \cite{leonski2001finite}. The resultant state  $\hat D(\beta)\hat U_{K}(\chi)\ket{\alpha}$ is amplitude-squeezed  \cite{Kitagawa1986PRA} and exhibits sub-Poissonian photon statistics $\langle \Delta\hat{n}^2\rangle<\langle \hat{n}\rangle$. The latter results in a zero-time-delay second-order coherence function $g^{(2)}(0)$ becoming less than unity, a nonclassical feature. 



In this work, we show that a number state $\vert N\rangle$ is realizable in a multi-step process.   Each step involves the application of the combined unitary operator  $\hat U(\beta,\chi)$ on a coherent state $\ket{\alpha}$,  with appropriate choices of $\beta$'s and $\chi$'s for each step. To produce $\ket{N}$ with high fidelity, we choose the amplitude of the coherent state as $\vert\alpha\vert\approx\sqrt{N}$.   The repeated operation of  $\hat D(\beta)\hat U_{K}(\chi)$ on an initial coherent state leads to a gradual narrowing of the photon number distribution and peaks at the mean photon number of the initial coherent state. This scheme allows high-fidelity generation of higher Fock states in circuit QED platforms. With the currently available experimental techniques, this scheme can also be implemented in a high-$Q$ cavity containing a Kerr nonlinear medium driven by a series of ultrashort coherent pulses. We note that there is no need for any photon blockade mechanism, unlike in the scheme presented in \cite{Lingenfelter2021ScAdv}. 

\begin{figure*}
\begin{center}
\includegraphics[scale=0.5]{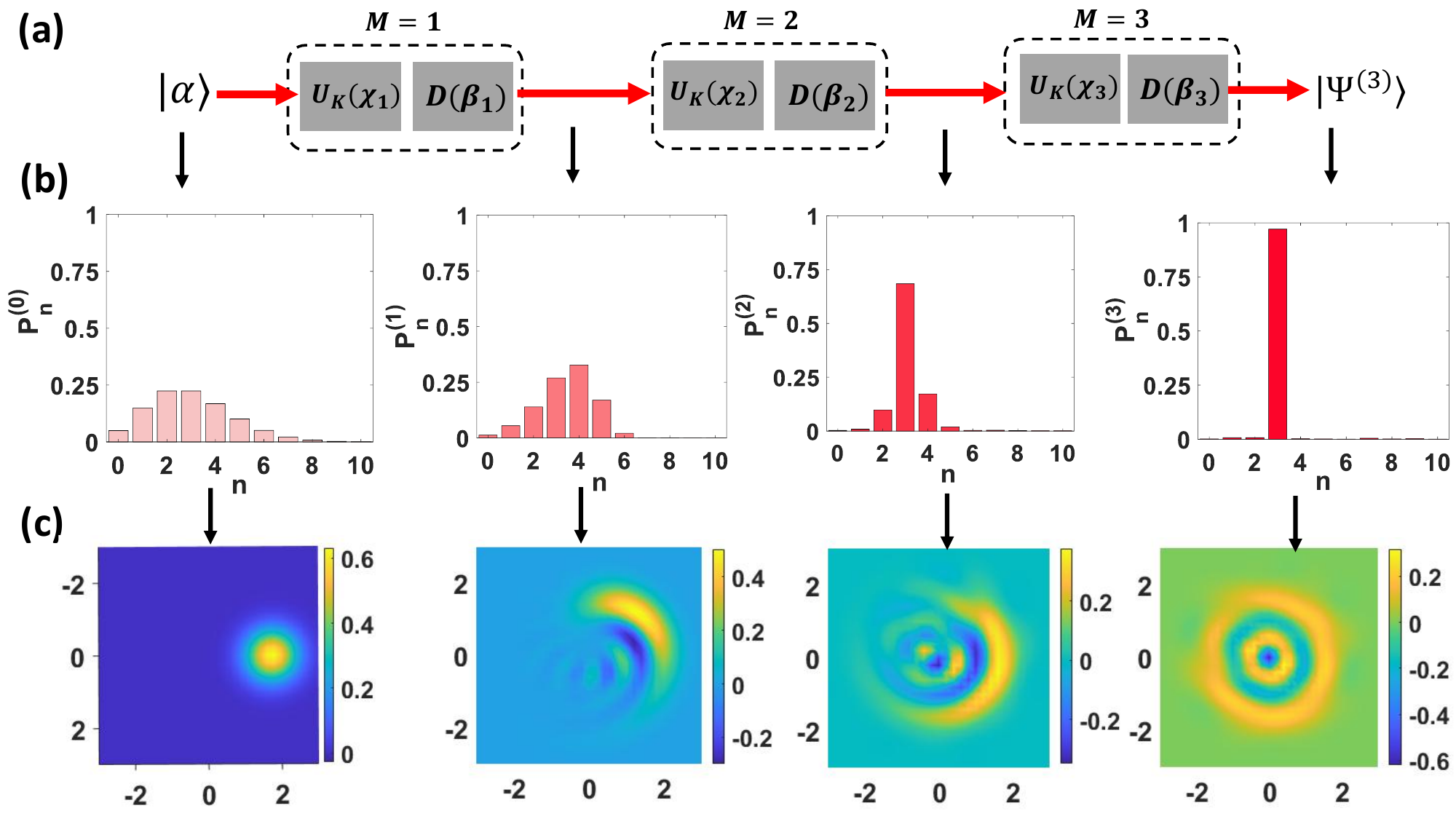}
\end{center}
\caption{ (a) Description of the repeated application of $\hat D(\beta_k)\hat U_{K}(\chi_k)$, up to $M=3$, applied on a coherent state $\ket{\alpha}$. (b) The probability distribution $P_n^{(M)}$ (fidelity) for $M=0,1,2$ and $3$ with $\alpha=\sqrt{3}$. After the third operation $(M=3)$, the resultant state has a fidelity of 0.97 with the Fock state $\ket{3}$. (c) Wigner functions of the resultant states after each operation.  For $M=3$, the Wigner function displays a ring-like structure with two rings due to its oscillations in the phase space, a characteristic of the Wigner function of the Fock state $\ket{3}$.   }\label{Scheme}
\end{figure*}

\begin{table*}[t]
\centering
\renewcommand{\arraystretch}{1.2}
\setlength{\tabcolsep}{6pt}

\begin{tabular}{|c|c|c|c|c|c|}
\hline
\multicolumn{6}{|c|}{${M = 2}$} \\ \hline
Fock state $(N)$ & $\beta_1$ & $\beta_2$ & $\chi_1/\pi$ & $\chi_2/\pi$ & Fidelity $(P_N^M)$ \\ \hline
1 & 0.31 & 0.41 & 0.74 & 0.7 & 0.97\\
2 & 0.45 & $-0.27$ & 0.85 & 0.63  & 0.93 \\
3 & 0.54 & 0.24  & 0.9 & 0.59  & 0.94 \\
4  & 0.64 & $-0.22$ & 0.927 & 0.57  & 0.94 \\
5 & 0.72 & 0.21 & 0.943 & 0.556  & 0.93 \\
6 & 0.77 & $-0.19$ & 1.047 & 0.453 & 0.91 \\
\hline
\end{tabular}


\begin{tabular}{|c|c|c|c|c|c|c|c|}
\hline
\multicolumn{8}{|c|}{${M = 3}$} \\ \hline
Fock state $(N)$ & $\beta_1$ & $\beta_2$ & $\beta_3$ &
$\chi_1/\pi$ & $\chi_2/\pi$ & $\chi_3/\pi$ & Fidelity  $(P_N^M)$ \\ \hline
1 & $-0.2$ &0.5 & 0.3 &  1 & 0.5 & 0.5 & 0.98 \\
2 & 0.26 & 0.45 & $-0.25$ & 1.915 & 0.22 & 1.34 & 0.97 \\
3 & 0.35 & 0.4 & 0.22 & 0.934 & 0.942 & 0.613 & 0.97 \\
4 & 0.52 & $-0.18$ & 0.2 & 1.936 & 1.27 & 0.41 & 0.97 \\
5  & 0.79 & 0.2 & 0.23 & 0.0647 & 0.223 & 1.466 & 0.97 \\
6 & 0.68 & 0.2 & 0.2 & 1.946 & 1.28 & 0.577 & 0.95\\
7 & 0.79 & 0.1 & 0.15 & 1.0415 & 0.785 & 1.673 & 0.95 \\
8 & 0.77 & $-0.13$ & 0.12 & 1.965 & 1.263 & 0.337 & 0.95 \\
9 & 0.64 & $-0.3$ & $-0.14$ & 0.972 & 1.934 & 0.65 & 0.95 \\
10 & 0.59 & $-0.31$ & 0.13 & 0.9743 & 1.9424 & 0.5296 & 0.96 \\
15 & 1.64 & 0.39 & $-0.11$ & 0.0187 & 1.967 & 0.4835 & 0.95 \\
20 & 0.95 & 0.23 & $-0.1$ & 0.013 & 0.873 & 1.5655 & 0.95 \\ \hline
\end{tabular}
 \caption{The optimal values of $\beta_k$ and $\chi_k$ to generate $N$-photon state with high fidelity after $M=2$ (top) and 3 (bottom) iterations. The value of the initial coherent state amplitude is taken to be $|\alpha|=\sqrt{N}$.}
 \label{ExpParameter}
\end{table*}

\section{Generation of Fock state}

The initial coherent state is a superposition of all the Fock states. To generate a Fock state, it is required to evolve the coherent state so that the fidelity of the chosen Fock state is increased to almost unity.   We consider two unitary evolutions, one corresponding to Kerr interaction with an effective Hamiltonian $\chi\hat{a}^{\dagger 2}\hat{a}^2$ and another to a displacement effected by $\beta\hat{a}^\dagger-\beta^*\hat{a}$.
The Kerr evolution operator commutes with the number operator  $\hat n=\hat a^\dagger \hat a$.  Therefore,   
the action of the Kerr unitary operator $\hat U_{K}(\chi)=e^{-i\chi \hat a^{\dagger 2} a^2}$ on a coherent state $\ket{\alpha}$  does not change the photon number distribution.  Consequently, the action of  $U_{K}(\chi)$ 
preserves the average number of photons $\langle\hat{n}\rangle=|\alpha|^2$ and photon number fluctuation $\Delta \hat n$ \cite{Yurke1986PRL}.  However, the phase-space distribution of the resultant state is non-Gaussian  and smeared \cite{Milburn1986PRA}.   If  $\chi=\pi/2$, the resultant state is a superposition of two coherent states $\ket{i\alpha}$ and $\ket{-i\alpha}$ is the Yurke-Stoler state \cite{Yurke1986PRL}. On the other hand, the action of a displacement operator $\hat D(\beta)=e^{\beta \hat a^\dagger-\beta^* \hat a}$ on the coherent state  $\ket{\alpha}$ results in another coherent state displaced  from $\vert\alpha\rangle$.    Under this transformation,  the average photon number changes while still preserving the Poissonian photon number distribution. However, the action of $\hat U_{K}(\chi)$ followed by $\hat D(\beta)$ on a coherent state does not preserve the average photon number and the photon-number fluctuations.   

Many interesting nonclassical states can be generated by choosing the values of $\chi$ and $\beta$ suitably.   The discussion in the following investigates repeated operations of combined Kerr evolution and displacement in a coherent state $\ket{\alpha}$.   Let  $M$ be the number of times  $\hat D(\beta_k) e^{-i\chi_k \hat a^{\dagger 2} \hat a^2}$ is operated on the coherent state $\ket{\alpha}$ [see Fig. \ref{Scheme}(a)]. 
The resultant state after $M$ such iterations is 
\begin{align}\label{Mthoperation}
\ket{\Psi^{(M)}}=\prod_{k=1}^M \left[\hat D(\beta_k) e^{-i\chi_k \hat a^{\dagger 2} \hat a^2}\right] \ket{\alpha},
\end{align} 
where $\hat D(\beta_k)$ is a displacement operator with amplitude $\beta_k$, and  $\chi_k$ is the Kerr strength.

For $M=1$,  the resultant state is $\ket{\Psi^{(1)}}= \hat D(\beta_1) e^{-i\chi_1 \hat a^{\dagger 2} \hat a^2} \ket{\alpha}$.  The probability of detecting $n$ photons in $\ket{\Psi^{(1)}}$, which is the fidelity of the Fock state $\ket{n}$, is given by
\begin{align}\label{ProbabilityDist1}
P_n^{(1)}=\left|e^{\frac{-|\alpha|^2}{2}}\sum_{m=0}^\infty \frac{\alpha^m e^{-i\chi_1 m(m-1)}}{\sqrt{m!}} \mathcal{D}_{n}^{m}(\beta_1) \right|^2,
\end{align}
where $\mathcal{D}_{n}^{m}(\beta_1)= \langle n | \hat D(\beta_1)|m\rangle$ are the matrix elements of the displacement operator \cite{AgarwalBook}, given by 
\begin{align}\label{MatrixElementDbeta}
&\bra{n}\hat D(\beta_1)\ket{m}\nonumber\\
&=e^{\frac{-|\beta_1|^2}{2}} \sqrt{\frac{m!}{n!}} (\beta_1)^{n-m} L_m^{(n-m)}(|\beta_1|^2) , ~~n\geq m \nonumber\\
&=e^{\frac{-|\beta_1|^2}{2}} \sqrt{\frac{n!}{m!}} (-\beta_1^*)^{m-n} L_n^{(m-n)}(|\beta_1|^2), ~~n\leq m.
\end{align}
Here,  $L_p^{q}(x)$ is the  the associated Laguerre polynomial \citep{AgarwalBook}. To detect a Fock state $|N\rangle$ in the resultant state $\ket{\Psi^{(1)}}$ with high fidelity, the values of $\alpha,\beta_1$ and $\chi_1$ must be appropriately chosen such that $P_N^{(1)}$ is maximum and is nearly 1. 

For a given $N$, we investigate Eq. \eqref{ProbabilityDist1} numerically to determine the optimal values for $\alpha,\beta_1$ and $\chi_1$ to obtain $\vert N\rangle$ with high fidelity.  We find that for $\alpha=\sqrt{1}, \beta_1= 0.5, $ and $\chi_1/\pi=1.61$, the fidelity of detecting the Fock state $\ket{1}$ compared to other Fock states in the resultant state $\ket{\Psi^{(1)}}$ reaches a maximum of $\sim 0.8$. Similarly, if $\alpha=\sqrt{2}$, the maximum fidelity of detecting two photons in the state $\ket{\Psi^{(1)}}$ is $\sim 0.73$ for $\beta_1=0.55$ and $\chi_1/\pi=1.835$.  It should be noted that, to maximize the fidelity of detecting the Fock state $\ket{N}$ in the resulting state, the amplitude of the initial coherent state should be chosen as $|\alpha|\approx \sqrt{N}$. However,  in a single operation ($M=1$), it is not possible to achieve Fock state fidelities close to unity for any value of $\beta_1$ and $\chi_1$. 

Now, consider $M>1$. By expressing Eq. \eqref{Mthoperation} in the Fock basis (see Appendix \ref{AppexState}), we find the fidelity of the target Fock state $\ket{N}$ in the resultant state $\ket{\Psi^{(M)}}$ for $M=2$ and 3 to be 
\begin{widetext}
\begin{subequations}\label{PN3}
\begin{align}
 P_N^{(2)}&=|\langle N\ket{\Psi^{(2)}}|^2=\left| e^{-|\alpha|^2/2} \sum_{{n}=0}^\infty \frac{\alpha^{n}}{\sqrt{{n}!}} e^{-i\chi_{1} n(n-1)} \sum_{r=0}^\infty  \mathcal{D}_{N}^{r}(\beta_2) \mathcal{D}_{r}^{n}(\beta_1) e^{-i\chi_{2} r(r-1)}\right|^2,\\
 P_N^{(3)}&=|\langle N\ket{\Psi^{(3)}}|^2=\left| e^{-|\alpha|^2/2} \sum_{{n}=0}^\infty \frac{\alpha^{n}}{\sqrt{{n}!}} e^{-i\chi_{1} n(n-1)} \sum_{r=0}^\infty \sum_{s=0}^\infty \mathcal{D}_{N}^{r}(\beta_3) \mathcal{D}_{r}^{s}(\beta_2) \mathcal{D}_{s}^{n}(\beta_1) e^{-i\chi_{3} r(r-1)} e^{-i\chi_{2} s(s-1)}\right|^2,
 \end{align}
 \end{subequations}
\end{widetext}
respectively.  In the above equation,  $\mathcal{D}_{n}^{m}(\beta_k)$ are the matrix elements of the displacement operator, and are defined in Eq. \eqref{MatrixElementDbeta}. The resultant state $\ket{\Psi^{(M)}}$ closely resembles the Fock state $\ket{N}$ if $P_{N}^{(M)}\approx 1$ for suitable values of $\alpha,\beta_k$ and $\chi_k$.  We set the initial coherent state amplitude to be $\alpha=\sqrt{N}$ and optimize the above equations numerically to find the optimal values of $\beta_k$ and $\chi_k$ by truncating the sums to a large cutoff ensuring the numerical convergence. The optimal values of $\beta_k$ and $\chi_k$ that maximize the fidelity of detecting a particular Fock state in the state $\ket{\Psi^{(M)}}$ given in Table.\ref{ExpParameter} are obtained by doing a grid search over  $\beta_k$ and $\chi_k$.  It should be noted that on applying $\hat U(\beta,\chi)$ twice, which corresponds to $M=2$, we can generate Fock states up to $N=6$ with fidelity more than $0.9$. For $N>6$, the fidelity of generating a Fock state falls below 0.9; hence, the optimized values of $\beta_k$ and $\chi_k$ for $M=2$ and $N>6$ are not listed in Table. \ref{ExpParameter}. However, these probabilities can be further improved if $M$ is higher, that is,  more iterations.  For example,   choosing $M=3$, we can generate a Fock state up to $N=20$ with a fidelity greater than $0.95$ (see Table. \ref{ExpParameter}).   The driving amplitudes ($\beta_k$) are chosen to be real, and we note that there is no significant improvement in fidelity when complex amplitudes are used instead. 

Transformations of the probability distribution and Wigner function of the coherent state to that of a Fock state (here $\ket{3}$) after each iteration are shown in Fig. \ref{Scheme}. The figure shows that after three steps, i.e.,  $M=3$, the probability distribution is concentrated around the Fock state $\ket{3}$, and those of the other Fock states in the superposition are negligible. Consequently, the Wigner function of the resultant state displays ring-like structures similar to the structures seen in the Wigner function of the Fock state $\ket{3}$. The values of $\beta_k$ and $\chi_k$ are considered from Table. \ref{ExpParameter} and the initial coherent state amplitude is taken as $\alpha=\sqrt{3}$. Hence, by adjusting the displacement amplitudes $\beta_k$ and Kerr strengths $\chi_k$, it is possible to modify the photon distribution of the initial coherent state to concentrate the probability distribution to a specific photon number during evolution.


\begin{figure*}
\begin{center}
\includegraphics[scale=0.5]{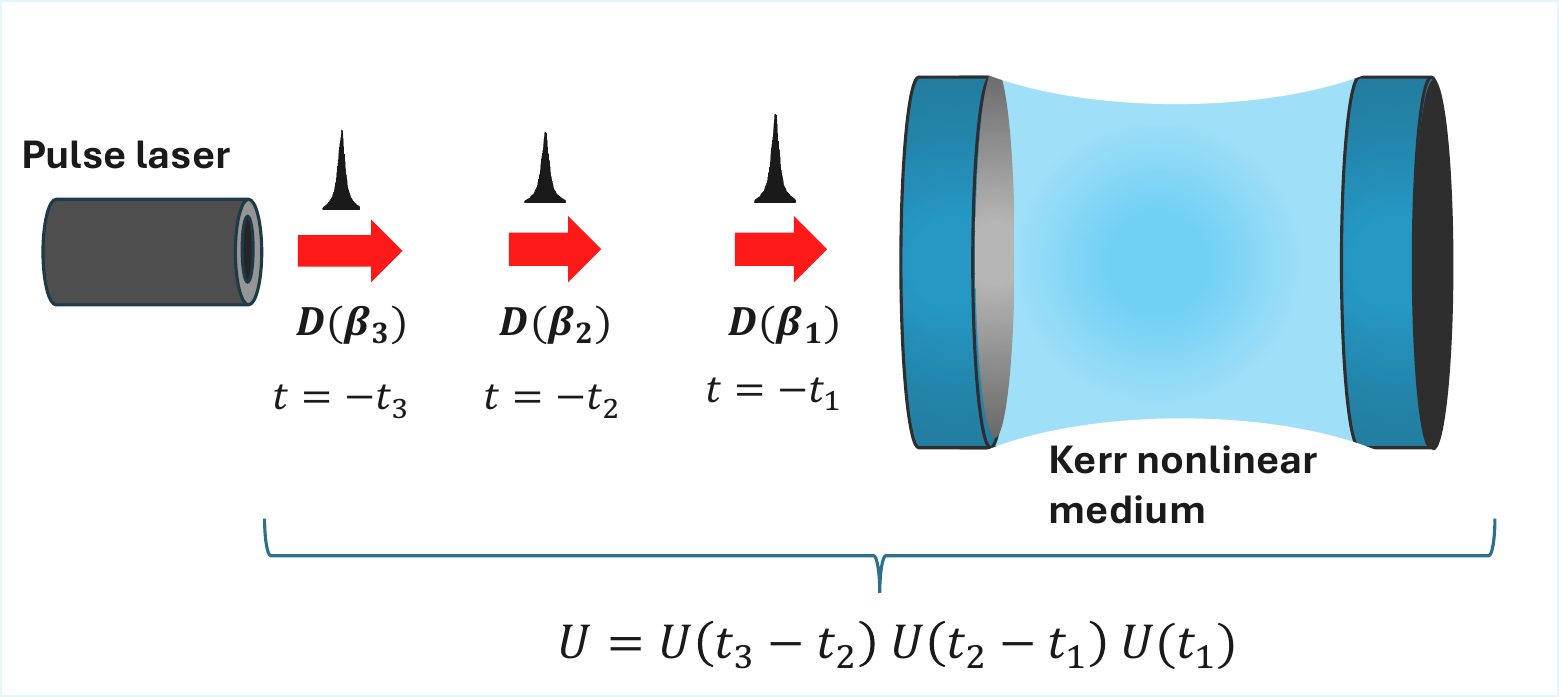}
\end{center}
\caption{ Schematic of a cavity filled with Kerr medium with Kerr strength $\mathcal{K}$. The cavity is driven by a series of ultra-short coherent pulses at $t=-t_1$, $t=-t_2$, and $t=-t_3$ such that they drive the cavity at time $t_1$, $t_2$, and $t_3$ respectively. The setup is equivalent to the unitary transformation of Eq. \eqref{Mthoperation}.   }\label{Experiment}
\end{figure*}

\section{Experimental feasibility}

The proposed scheme involves a cavity filled with a Kerr medium driven by a series of ultra-short coherent pulses, each pulse in resonance with the cavity mode frequency (Fig. \ref{Experiment}). These pulses are represented by sufficiently narrow functions. The Hamiltonian 
\begin{equation}\label{KerrCavityHamiltonian}
\hat{H}=\mathcal{K} \hat a^{\dagger 2} \hat a^2+i f(t-t_1)\beta_1( \hat a^\dagger- \hat a), 
\end{equation}
describes a cavity containing Kerr nonlinearity with strength $\mathcal{K}$, excited by a coherent pulse at $t= t_1$. The function $f(t-t_1)$ is considered to be a quasi-delta function peaked at $t_1$.  In the limit $f(t-t_1)\rightarrow \delta(t-t_1)$, we have the unitary evolution operator at time $t_1$ for the system to be (see Appendix. \ref{ApexDispKerr}) 
\begin{align}\label{DispKerrUnitary}
\hat U(t_1)=\hat D(\beta_1)e^{-i\mathcal{K} t_1 \hat a^{\dagger 2}\hat a^2},
\end{align}
where $ \hat D(\beta_1)=\text{exp}\left[{\beta_1 (\hat a^\dagger-\hat a)}\right]$, is the displacement operator $\hat D( \beta_1)$. Here, $\beta_1$ is the amplitude (assumed to be real)  of the first pulse, and $\mathcal{K} t_1$ is the time of evolution of the cavity field in the presence of the Kerr medium. The evolution of the initial state from the instant right after the first pulse and up to the second pulse at time $t_2$, i.e., during the interval $t_2-t_1$, is governed by the unitary operator $\hat U(t_2-t_1)=\hat D(\beta_2)e^{-i\mathcal{K} (t_2-t_1) \hat a^{\dagger 2}\hat a^2}$, where $\beta_2$ is the amplitude of the second pulse.
Similarly,  for the duration $t_3-t_2$,  the relevant unitary evolution operator is $\hat U(t_3-t_2)=\hat D(\beta_3)e^{-i\mathcal{K} (t_3-t_2) \hat a^{\dagger 2}\hat a^2}$, where $\beta_3$ is the amplitude of the third pulse. 
It is to be noted that the normalized time $\mathcal{K} t_1$, $\mathcal{K} (t_2-t_1)$ and $\mathcal{K} (t_3-t_2)$ are respectively analogous to $\chi_1$, $\chi_2$ and $\chi_3$ in Eq. \eqref{Mthoperation}. In short, the nonlinear interaction term involves both the nonlinearity strength and duration of evolution,  while the displacement term involves only the amplitude of displacement and not the evolution duration. Therefore, the nonlinear parameters $\chi_1,\chi_2$ and $\chi_3$ of Eq. \eqref{Mthoperation} can be tuned by controlling the time intervals between the pulses for a given Kerr strength $\mathcal{K}$. 

\begin{figure*}
\begin{center}
\includegraphics[scale=0.40]{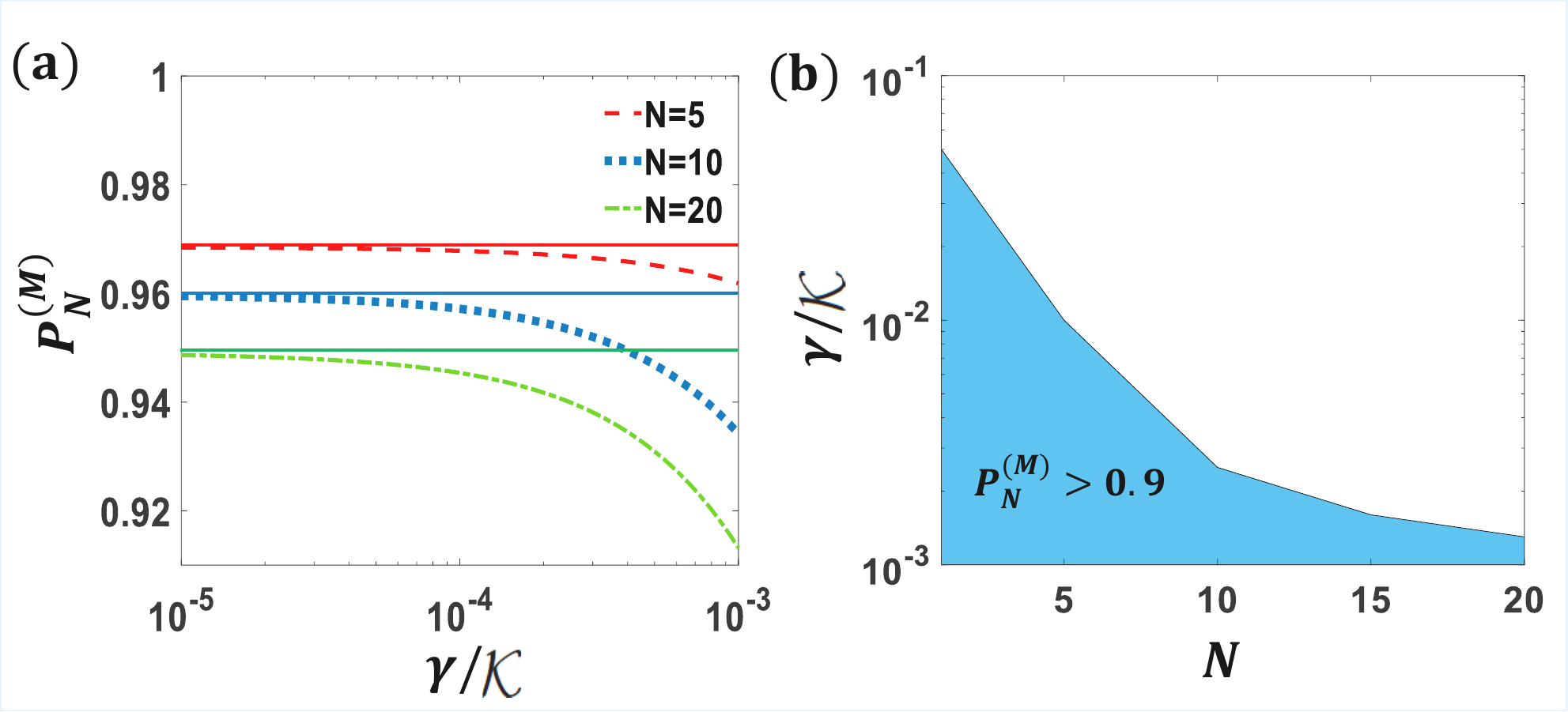}
\end{center}
\caption{ (a) Fidelity $(P_N^{(M)})$ of generated states with Fock states for $N=5,10,20$ and  $M=3$ as a function of cavity decay rate, normalized to Kerr strength. The horizontal lines correspond to the fidelities in the absence of dissipation.  (b) Performance of the protocol  shown as a shaded region in the $\gamma/\mathcal{K}$ versus $N$ plot. The shaded region corresponds to the fidelity larger than 0.9, indicating for a given $N$, the range of $\gamma/\mathcal{K}$ for which $P_N^M>0.9$.   }\label{PhotonLoss}
\end{figure*}

In cavity QED experiments, photon loss from the cavity to the environment is unavoidable. The effect of photon loss on the preparation of the Fock state is analyzed using the following master equation
\begin{align}
\frac{\partial \rho}{\partial t}=-i[\hat H,\rho]+\frac{\gamma}{2}(2\hat a\rho \hat a^\dagger-\hat a^\dagger \hat a\rho-\rho \hat a^\dagger \hat a),
\end{align}
where $\gamma$ is the decay rate of the cavity and $\hat H=\mathcal{K} \hat a^{\dagger 2} \hat a^2+i\sum_j \delta(t-t_j)\beta_j( \hat a^\dagger- \hat a)$ is the Hamiltonian for a Kerr cavity driven by a series of ultra-short pulses \cite{milburn1991quantum}. Although we discussed the possible implementation of the required unitary transformation for Fock state generation in a cavity QED system,  the Hamiltonian given in Eq. \eqref{KerrCavityHamiltonian} also describes anharmonic, driven superconducting resonators in circuit QED systems \cite{gevorgyan2012generation}. Therefore, the method described for generating large Fock states is also applicable to the circuit QED platform.  With the currently achievable experimental values of Kerr nonlinear strength $\mathcal{K}/2\pi=12.5$ MHz and cavity decay rate $\gamma/\mathcal{K} \sim 10^{-5}-10^{-3}$ in circuit QED system \cite{Vrajitoarea2020NatPhys, Hajr2024PRX},  numerical investigations of the master equation show that it is possible to achieve a fidelity more than 0.9 in generating Fock states with quantum numbers up to $N=20$ as shown in  Fig. \ref{PhotonLoss}(a).

The output fidelity is a good measure of the performance.   We set a fidelity threshold of 0.9 to indicate good performance.  Fig. \ref{PhotonLoss}(b)  shows the region in the $\gamma/\mathcal{K}$  versus $N$  space where the protocol exhibits good performance.  The graph can be interpreted to mean that performance degrades with increasing photon loss, as expected, or that it requires higher Kerr strength to compensate for the loss of fidelity due to photon loss. In the absence of dissipation $(\gamma=0)$, fidelity improves with the number of iterations $(M)$. However, if $\gamma\neq 0$,  increasing the number of iterations requires a longer field evolution duration $(t_M)$, which will degrade fidelity due to photon loss from the cavity.



\section{Summary}
We presented a novel quantum state engineering scheme to generate Fock states via repeated Kerr and displacement operations on a suitable coherent state.  The scheme is realizable in a cavity containing a Kerr nonlinear medium and driven by a series of ultra-short laser pulses.   With the currently achievable experimental parameters, generating a large Fock state with fidelity greater than 0.9 is possible even in the presence of substantial cavity loss when the cavity damping rate is as high as one-thousandth of the nonlinear strength.

We have verified that fidelity above 0.9 is achievable for Fock states with up to 20 photons, even with the number of iterations limited to three. Increasing the number of iterations beyond three will improve fidelity in the generation of Fock states, even with photon numbers more than twenty. Our discussions on the generation of the Fock state in cavities are also applicable to circuit-QED platforms \cite{touzard2019gated,rebic2009giant} and optomechanical systems \cite{li2025kerr}, where displacement operations and Kerr nonlinearities are realizable. 

The unitary operator given in Eq. \eqref{DispKerrUnitary} is derived assuming the driving to be a delta-function.   When the pulse width is non-zero, and the loss-to-nonlinearity ratio ($\gamma/\mathcal{K}$) is not small, we must retain higher-order terms in the Magnus expansion \cite{blanes2009magnus,horoshko2021time}.   The effect of time-ordering \cite{quesada2015time}  combined with a higher $\gamma/\mathcal{K}$  ratio on the generation of Fock states will be investigated in a future study.   The other question of interest is the use of squeezing operations in this scheme. We plan to analyze schemes that use Kerr nonlinearity,  displacement and squeezing to generate nonclassical states such as superpositions of number states, and the Gottesman-Kitaev-Preskill (GKP) qubits \cite{gottesman2001encoding}. 

\noindent
\section*{Acknowledgements} A.P. thanks DST for their
funding support through the National Quantum Mission(NQM).

\section*{Data availability} 
The data that support the findings of this article are openly available at \cite{mydata2025}

\appendix
\begin{widetext}
\section{Calculation of the resultant states and probabilities}\label{AppexState}
For $M=1$, we have
\begin{align}
\ket{\Psi^{(1)}}&= \hat D(\beta_1) e^{-i\chi_1 \hat a^{\dagger 2} \hat a^2} \ket{\alpha}, \nonumber\\
&= \sum_{m_0=0}^\infty e^{-|\alpha|^2/2}\frac{\alpha^{m_0}}{\sqrt{{m_0}!}} \hat D(\beta_1) e^{-i\chi_1 \hat a^{\dagger 2} \hat a^2} \ket{{m_0}},\nonumber\\
&= \sum_{{m_0}=0}^\infty e^{-|\alpha|^2/2}\frac{\alpha^{m_0}}{\sqrt{{m_0}!}}\left(\sum_{m_1=0}^\infty \ket{m_1}\bra{m_1} \right) \hat D(\beta_1) e^{-i\chi_1 \hat a^{\dagger 2} \hat a^2} \ket{{m_0}},\nonumber\\
&= \sum_{{m_0}=0}^\infty e^{-|\alpha|^2/2}\frac{\alpha^{m_0}}{\sqrt{{m_0}!}}\sum_{m_1=0}^\infty \mathcal{D}_{m_1}^{m_0}(\beta_1) e^{-i\chi_1 m_0(m_0-1)}  \ket{m_1},
\end{align}
where $\mathcal{D}_{m_1}^{m_0}(\beta_1)=\bra{m_1} \hat D(\beta_1) \ket{m_0} $.

Similarly, for $M=2$, we have
\begin{align}
\ket{\Psi^{(2)}}&=  \hat D(\beta_2) e^{-i\chi_2 \hat a^{\dagger 2} \hat a^2} \hat D(\beta_1) e^{-i\chi_1 \hat a^{\dagger 2} \hat a^2} \ket{\alpha}, \nonumber\\
&= \sum_{{m_0}=0}^\infty e^{-|\alpha|^2/2}\frac{\alpha^{m_0}}{\sqrt{{m_0}!}}  \sum_{m_2=0}^\infty  \sum_{m_1=0}^\infty \mathcal{D}_{m_2}^{m_1}(\beta_1) e^{-i\chi_2 m_1(m_1-1)} \mathcal{D}_{m_1}^{m_0}(\beta_2) e^{-i\chi_1 m_0(m_0-1)} \ket{m_2}
\end{align}
where $\mathcal{D}_{m_2}^{m_1}(\beta_2)= \bra{m_2}\hat D(\beta_2)\ket{m_1}$ and $\mathcal{D}_{m_1}^{m_0}(\beta_1)= \bra{m_1}  \hat D(\beta_1) \ket{m_0}$.

In general, after $M$-th iteration
\begin{align}
\ket{\Psi^{(M)}}&= \sum_{{m_0}=0}^\infty e^{-|\alpha|^2/2}\frac{\alpha^{m_0}}{\sqrt{{m_0}!}}\prod_{k=1}^M\left[\sum_{m_k=0}^\infty \mathcal{D}_{m_k}^{m_{k-1}}(\beta_k) e^{-i\chi_{k} m_{k-1}(m_{k-1}-1)}\right] \ket{m_M},
\end{align}
where $\mathcal{D}_{m_{k}}^{m_{k-1}}(\beta_k)= \langle m_k | \hat D(\beta_k)|m_{k-1}\rangle$.

For $M=2$,  the probability (fidelity) of detecting $N$ photons is
\begin{align}
    P_N^{(2)}=|\langle N |\Psi^{(2)}\rangle |^2=e^{-|\alpha|^2} \left|  \sum_{{n}=0}^\infty \frac{\alpha^{n}}{\sqrt{{n}!}} e^{-i\chi_{1} n(n-1)} \sum_{r=0}^\infty  \mathcal{D}_{N}^{r}(\beta_2) \mathcal{D}_{r}^{n}(\beta_1) e^{-i\chi_{2} r(r-1)}\right|^2,
\end{align}
and for $M=3$, 
\begin{align}
    P_N^{(3)}=|\langle N |\Psi^{(3)}\rangle |^2=e^{-|\alpha|^2} \left| \sum_{{n}=0}^\infty \frac{\alpha^{n}}{\sqrt{{n}!}} e^{-i\chi_{1} n(n-1)} \sum_{r=0}^\infty \sum_{s=0}^\infty \mathcal{D}_{N}^{r}(\beta_3) \mathcal{D}_{r}^{s}(\beta_2) \mathcal{D}_{s}^{n}(\beta_1) e^{-i\chi_{3} r(r-1)} e^{-i\chi_{2} s(s-1)}\right|^2.
\end{align}
\vspace{1cm}


\section{Calculation of the unitary operator for a Kerr cavity driven by ultra-short pulses}\label{ApexDispKerr}
Consider a Kerr cavity driven by an ultra-short pulse. The Hamiltonian is 
\begin{equation}\label{fullH}
\hat{H}=\hat H_0+\hat H_I(t)=\mathcal{K} \hat a^{\dagger 2} \hat a^2+i f(t-t_1)\beta_1( \hat a^\dagger- \hat a), 
\end{equation}
where $f(t-t_1)$ is a sharply peaked function centered at $t_1$, $\hat H_0=\mathcal{K} \hat a^{\dagger 2} \hat a^2$ and $\hat H_I(t)=i f(t-t_1)\beta_1( \hat a^\dagger- \hat a)$.

Define the first interaction picture evolution operator \cite{horoshko2021time},
\begin{align}\label{InteractionUnitary}
\hat U_I(t)=\hat U_0^\dagger(t) \hat U(t),
\end{align}
{where $\hat U_0(t)=e^{-i\int_0^t \hat H_0dt'}=e^{-i \hat H_0 t}$ and the time-ordered unitary operator $\hat U(t)=\mathcal{T} e^{-i\int_0^t \hat H(t')dt'}$ corresponds to evolution under the full Hamiltonian $\hat H(t)$ given in Eq. \eqref{fullH}, $\mathcal{T}$ being the time-ordering operator}.    This operator satisfies  
\begin{align}\label{EvolutionofUI}
\frac{d}{dt} \hat U_I(t)=-i \hat{\tilde{H}}_I(t) \hat U_I(t),
\end{align}
where $\hat{\tilde{H}}_I(t)=\hat U_0^\dagger \hat H_I(t) \hat U_0=if(t-t_1)\beta_1(\hat A^\dagger(t)-\hat A(t))$ with $\hat A(t)= e^{-2i\mathcal{K} t\hat a^\dagger \hat a}\hat a$. The solution of Eq. \eqref{EvolutionofUI} is
\begin{align} \label{UnitaryInteraction}
\hat U_I(t)=\mathcal{T}\exp\left(-i\int_0^t \hat{\tilde{H}}_I(t')dt'\right). 
\end{align}
Using Magnus expansion \cite{blanes2009magnus}, the time-ordered unitary operator $\hat U_I(t)$ can be expanded as
\begin{align}
\hat U_I(t)=e^{\Omega_1+\Omega_2+\Omega_3+...}
\end{align}
where the first three exponents are
\begin{align}
\Omega_1&=-i\int_0^t dt'_1\hat{\tilde{H}}_I(t'_1),\nonumber\\
\Omega_2&=-\frac{1}{2}\int_0^t dt'_1\int_0^{t'_1}dt'_2[\hat{\tilde{H}}_I(t'_1),\hat{\tilde{H}}_I(t'_2)],\nonumber\\
\Omega_3&=\frac{i}{6}\int_0^t dt'_1\int_0^{t'_1}dt'_2\int_0^{t'_2}dt'_3([\hat{\tilde{H}}_I(t'_1),[ \hat{\tilde{H}}_I(t'_2),\hat{\tilde{H}}_I(t'_3)]]+[[\hat{\tilde{H}}_I(t'_1),\hat{\tilde{H}}_I(t'_2)],\hat{\tilde{H}}_I(t'_3)]).
\end{align}

If  $f(t)=\delta(t-t_1)$,  then 
\begin{align}
\Omega_1&=-i\int_0^t dt'_1 i\beta_1 f(t'_1-t_1)(\hat A^\dagger(t'_1)-\hat A(t'_1)), \nonumber\\
&=\beta_1 \int_0^t dt'_1 \delta(t'_1-t_1)(\hat A^\dagger(t'_1)-\hat A(t'_1))=\beta_1(\hat A^\dagger (t_1)-\hat A(t_1)).
\end{align}  
The second exponent is 
\begin{align}
\Omega_2&=-\frac{1}{2}\int_0^t dt'_1\int_0^{t'_1}dt'_2[\hat{\tilde{H}}_I(t'_1),\hat{\tilde{H}}_I(t'_2)],\nonumber\\
&=-\frac{1}{2}\int_0^t dt'_1\int_0^{t'_1}dt'_2 i^2\beta_1^2 f(t'_1-t_1)f(t'_2-t_1)[\hat A^\dagger (t'_1)-\hat A(t'_1),\hat A^\dagger (t'_2)-\hat A(t'_2)],\nonumber\\
&=\frac{\beta_1^2}{2}\int_0^t dt'_1\int_0^{t'_1}dt'_2  \delta(t'_1-t_1)\delta(t'_2-t_1)[\hat A^\dagger (t'_1)-\hat A(t'_1),\hat A^\dagger (t'_2)-\hat A(t'_2)],\nonumber\\
&=\frac{\beta_1^2}{2}[\hat A^\dagger (t_1)-\hat A(t_1),\hat A^\dagger (t_1)-\hat A(t_1)]=0.
\end{align}
Similarly,  the higher-order terms also vanish in the delta-function limit. Therefore, the unitary operator given in Eq. \eqref{UnitaryInteraction} becomes
\begin{align}
\hat U_I(t_1)=e^{\beta_1(\hat A^\dagger (t_1)-\hat A(t_1))}.
\end{align}
Substituting this result in Eq. \eqref{InteractionUnitary}, we get the operator at time $t_1$ to be
\begin{align}\label{DbetaUchiAppendix}
\hat U(t_1)=\hat U_0(t_1)\hat U_I(t_1)&=\hat U_0 (t_1) \hat U_I(t_1)\hat U_0^\dagger(t_1) \hat U_0(t_1)\nonumber\\
&=e^{\beta_1(\hat a^\dagger-\hat a)}e^{-i\mathcal{K}t_1\hat a^{\dagger 2}\hat a^2} \nonumber\\
&=\hat D(\beta_1)e^{-i\mathcal{K} t_1\hat a^{\dagger 2}\hat a^2}, 
\end{align}
where we used the transformation $\hat U_0 (t_1) \hat U_I(t_1)\hat U_0^\dagger(t_1)=\hat U_0 (t_1) e^{\beta_1(\hat A^\dagger (t_1)-\hat A(t_1))}\hat U_0^\dagger(t_1)=e^{\beta_1(\hat a^\dagger-\hat a)}$. 
Therefore, a Kerr cavity driven by an ultra-short pulse at $t_1$ can be described by the unitary operator $\hat D(\beta_1)e^{-i\mathcal{K} t_1\hat a^{\dagger 2}\hat a^2}$.


\end{widetext}
%

\end{document}